\documentclass[a4paper,twocolumn,english,aps,prl,superscriptaddress,twocolumn]{revtex4-2}
\usepackage[T1]{fontenc}
\usepackage[latin9]{inputenc}
\usepackage{color}
\usepackage{babel}
\usepackage{verbatim}
\usepackage{bm}
\usepackage{amsmath}
\usepackage{amssymb}
\usepackage{graphicx}
\usepackage[pdfusetitle,
 bookmarks=true,bookmarksnumbered=false,bookmarksopen=false,
 breaklinks=false,pdfborder={0 0 1},backref=false,colorlinks=false]
 {hyperref}

\makeatletter

\usepackage{babel}
\usepackage{babel}

\makeatother

\begin{document}
\renewcommand{\figurename}{FIG.}
\title{Phase Transitions in Disordered Altermagnetic Josephson Junctions}
\author{Chang-An Li}
\email{changanli@ustc.edu.cn}

\affiliation{Hefei National Laboratory, Hefei, Anhui 230088, China}
\affiliation{School of Emerging Technology, University of Science and Technology
of China, Hefei, Anhui 230026, China}
\affiliation{Institute for Theoretical Physics and Astrophysics, University of
W\"urzburg, 97074 W\"urzburg, Germany}
\date{\today}
\begin{abstract}
Altermagnetic Josephson junctions (AMJJs) can host unconventional
$\pi$ phase and $\varphi$ phase despite vanishing net magnetizations.
Whether these phases are stable against disorder existing in real
materials remains an open question. Here, we investigate impact of
disorder on exotic phases in two-dimensional AMJJs consisting of two
conventional superconductors mediated by a $d$-wave altermagnet.
We show that disorder is able to drive phase transitions from the
exotic $\pi$ phase to conventional $0$ phase, accompanied by a substantial
suppression of critical current. This behavior is attributed to modifications
of the tunneling Cooper-pair phase shift and superconducting decoherence.
Remarkably, the anomalous $\varphi$ phase is highly fragile in presence
of disorder and can be driven to either a $\pi$ phase or $0$ phase
in a nonreciprocal way. Across such transitions, the first harmonic
of current-phase relation changes its sign, while the higher-order
harmonics are rapidly suppressed. Our findings reveal the crucial
role of disorder in tailoring distinct phases of AMJJs and shed new
light on their potential functionalities.
\end{abstract}
\maketitle

\section{Introduction}

Altermagnets (AMs) have emerged as a new class of magnetic phase characterized
by momentum-dependent spin splitting but with vanishing net magnetization
\citep{Libor22prx1,Libor22prx2,BaiL24AFM,Song25NRM,Jungwirth25Newton}\textcolor{black}{.}
The AMs combine key features of conventional ferromagnets with spin-split
band structures and antiferromagnets with net zero magnetization,
thereby providing a fertile platform for novel physics \citep{Rafael21PRL,ShaoDF21NC,ZhangR24prl,Brekke23prb,LeeS24prl,Krempasky24Nature}
and potential applications \citep{Libor23PRL,ZYLiu24PRL,Ghorashi24PRL,ZHouXD24PRL,DuanXK25PRL,GuMQ25PRL,LiuL26PRL,LiCA26prl}.
In particular, the vanishing net magnetization eliminates the detrimental
stray fields when proximitized to superconductors (SCs) \citep{Blamire14JPCM,Giil24prb,Kirill25arxiv}.
Henceforth, the hybrid structure of AM-SC provides promising functionalities
\citep{SunC23prb,Papaj23PRB,Das24prb,Mondal25prb,Mazin25arXiv}. A
prominent example is the altermagnetic Josephson junction (AMJJ) \citep{Ouassou23PRL,Beenakker23prb,ZhangSB24NC,ChengQ24prb,Fukaya25prb,Fukaya25JPCM}
showing anomalous superconducting transport properties \citep{Chakraborty25prl,Sim05prb,ZhaoW25prb,SunHP25prb,LiC26prl}.
With $d$-wave AMs, the AMJJ demonstrates tunable $0$-$\pi$ phase
oscillations as a function of the junction length \citep{Ouassou23PRL,Beenakker23prb,ZhangSB24NC,ChengQ24prb},
analogous to those previously studied in ferromagnetic or antiferromagnetic
Josephson junctions with finite magnetizations \citep{Ryazanov01prl,Buzdin05rmp,Andersen06prl,Kawabata10prl}.
Such controllable $\pi$-phase Josephson junctions are of particular
interest for superconducting qubits and digital circuits \citep{Ioffe99Nature,Feofanov10NP,Testa04APL}.
In addition, AMJJs can host anomalous $\varphi$ phase with supercurrent
nodal points residing at general phase differences $\phi\neq0,\pi$
\citep{LuB24prl}. However, to construct AMJJs with the confirmed
AM materials, such as \textcolor{black}{$\mathrm{KV_{2}Se_{2}O}$}\ \textcolor{black}{\citep{Ma2021NC,JiangB25NP,FYZhang25NP}
and $\mathrm{CrSb}$}\ \textcolor{black}{\citep{Reimers24NC,DingJY24PRL,ZengM24AS,YangGW25NC},}
the presence of disorder will significantly modify their quantum transport
properties \citep{LiCA26prl}. Henceforth, a crucial open question
is how these exotic phases of AMJJs will be influenced by disorder?
Addressing this problem will not only reveal the unique interplay
between AMs, disorder, and superconductivity, but also provide useful
guidance for practical implementations of AM-based Josephson devices.

\begin{figure}
\includegraphics[width=1\linewidth]{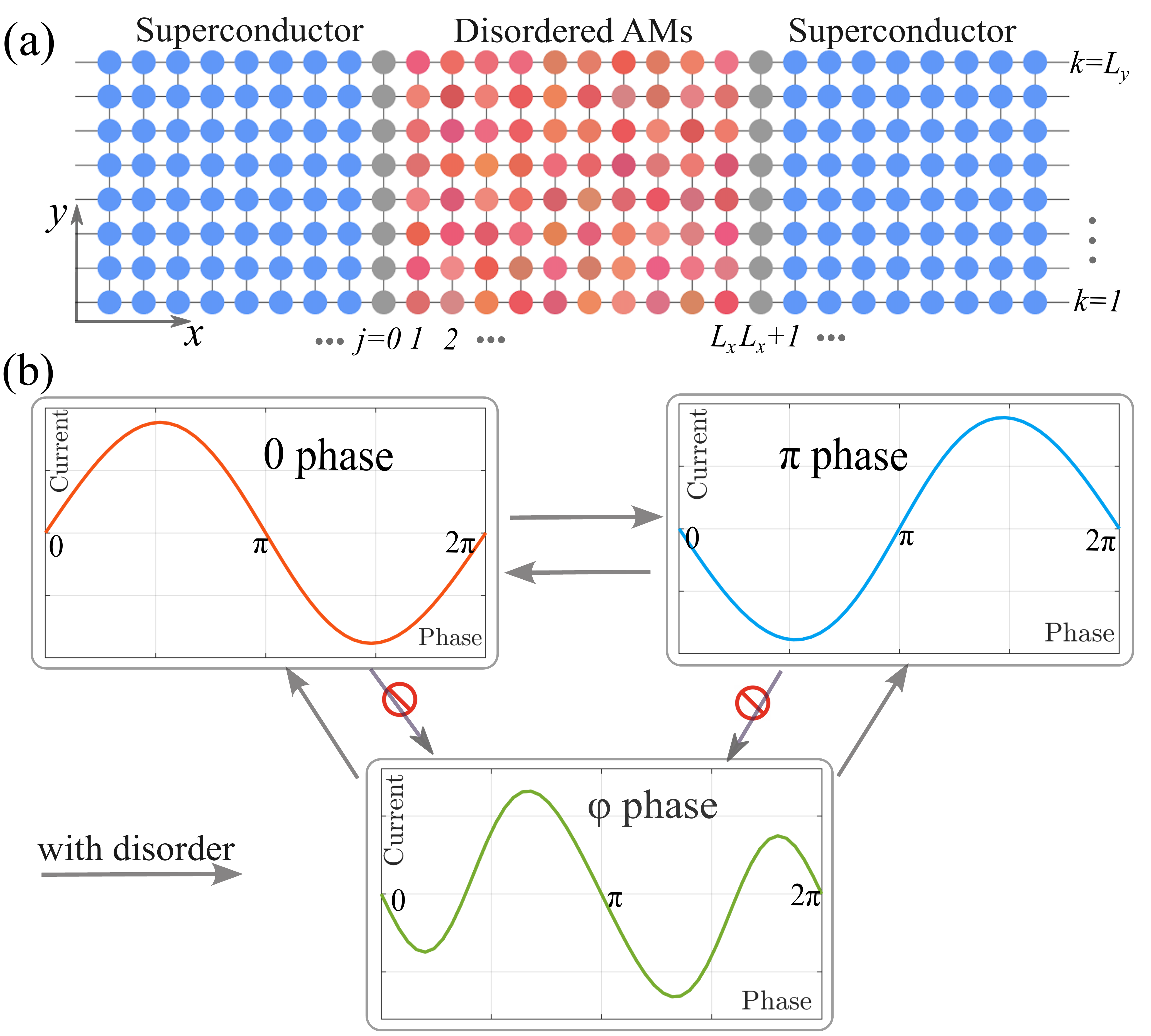}

\caption{(a) Schematic of the disordered AMJJ in 2D. Two $s$-wave superconductor
leads (blue color) with a phase difference $\phi$ are connected by
a disordered $d$-wave AM ($d_{x^{2}-y^{2}}$-wave or $d_{xy}$-wave)
with a length of $L_{x}$ (red color). The color variation indicates
random onsite potentials at different lattice sites. There are normal
metals at two interfaces with a barrier potential of $V_{b}$ (gray
color). Periodic boundary conditions are applied along $y$ direction.
(b) Sketch of the disorder-induced phase transitions between different
phases of the AMJJ. Note that the transitions between $0$ phase (red
color) and $\pi$ phase (blue color) are reciprocal, whereas the transition
from $\varphi$ phase (green color) to either $0$ phase or $\text{\ensuremath{\pi}}$
phase is nonreciprocal. \label{fig1:main-result}}
\end{figure}

In this work, we systematically investigate the impact of disorder
on exotic phases of AMJJs in two dimensions (2D). The AMJJ consists
of an SC-AM hybrid structure as shown in Fig. \ref{fig1:main-result}(a),
where disorder is introduced in the $d$-wave altermagnetic junction
region. We find that the exotic $\pi$ phase in AMJJs is sensitive
to disorder and can be driven to the conventional $0$ phase, even
at a relatively weak disorder strength. The corresponding critical
supercurrent will be strongly suppressed. Disorder can also induce
a transition from the conventional $0$ phase to the exotic $\pi$
phase in the reverse direction, making $0$-$\pi$ phase transition
reciprocally possible. Remarkably, the anomalous $\varphi$ phase
turns out to be highly fragile against disorder. It is eventually
driven to either a $\pi$ phase or $0$ phase in a non-reciprocal
manner. These results hold generally for different altermagnetic orientations
and junction details. We demonstrate strong numerical evidence for
these phenomena and provide direct interpretations based on disorder-induced
phase shift of Cooper pairs tunneling across the junction.

\section{Model and Methods}

We consider a disordered AMJJ schematically illustrated in Fig. \ref{fig1:main-result}(a).
The system consists of two superconductor leads ($j<0$ and $j>L_{x}+1$)
separated by a disordered $d$-wave AMs (i.e., $1\leq j\leq L_{x}$).
At the interfaces, thin normal-metal barrier layers are introduced
with a control potential $V_{b}$. The lattice sites can be labeled
by the basis vector $\bm{r}=j{\bf e}_{x}+k{\bf e}_{y}$, where ${\bf e}_{x}$
and ${\bf e}_{y}$ are unit vectors in $x$ and $y$ directions, respectively.
The lattice constant is set to be unit for simplicity. Periodic boundary
conditions are imposed along $y$ direction. The mean-field tight-binding
Hamiltonian reads
\begin{alignat}{1}
\mathcal{H}(\phi)= & \sum_{\bm{r},\bm{r}'}[\Psi^{\dagger}_{\bm{r}}h_{\bm{r},\bm{r}'}\Psi_{\bm{r}'}-\bar{\Psi}_{\bm{r}}h^{*}_{\bm{r},\bm{r}'}\bar{\Psi}^{\dagger}_{\bm{r}'}]\nonumber \\
 & +\sum_{\bm{r}\in\mathrm{SCs}}[\Psi^{\dagger}_{\bm{r}}\Delta_{\bm{r}}\bar{\Psi}^{\dagger}_{\bm{r}}-\bar{\Psi}_{\bm{r}}\Delta^{*}_{\bm{r}}\Psi_{\bm{r}}],
\end{alignat}
where $\Psi_{\bm{r}}=(C_{\bm{r}\uparrow},C_{\bm{r}\downarrow})^{T}$
with $C^{\dagger}_{\bm{r}\sigma}$ ($C_{\bm{r}\sigma}$) the creation
(annihilation) operator for an electron at site $\bm{r}$ of spin
$\sigma=\{\uparrow,\downarrow\}$. $\bar{\Psi}_{\bm{r}}$ means the
transpose of $\Psi_{\bm{r}}$. The spin-singlet pairing potential
takes $\Delta_{\bm{r}\in L}=i\sigma_{y}\Delta$ and $\Delta_{\bm{r}\in R}=i\sigma_{y}\Delta e^{i\phi}$
with a phase difference $\phi$ across the left (L) and the right
(R) lead. Explicitly, the normal-state Hamiltonian $h_{\bm{r},\bm{r}'}$
containing the $d_{x^{2}-y^{2}}$-wave AMs junction region on a square
lattice is given by \citep{Libor22prx1,Libor22prx2}

\begin{equation}
h_{\bm{r},\bm{r}'}=\sum_{\sigma,\sigma'}\{-t\delta_{\left|\bm{r}-\bm{r}'\right|,1}-t_{J,\bm{r}}[{\bf e}_{\langle\bm{r}\bm{r}'\rangle}\cdot\bm{\sigma}]_{\sigma\sigma'}-\mu)\},
\end{equation}
where $t$ is the nearest-neighbor hopping amplitude, $\mu$ is the
chemical potential, and $\bm{\sigma}=(\sigma_{x},\sigma_{y},\sigma_{z})$
denotes the Pauli matrix vector for spin degrees of freedom. The spin-dependent
altermagnetic hopping amplitude takes $t_{J,\bm{r}}=t_{J}$ in the
AM region and vanishes elsewhere. The unit vector ${\bf e}_{\langle\bm{r}\bm{r}'\rangle}$
for nearest-neighbor hopping in AMs is set as ${\bf e}_{\langle\bm{r}\bm{r}'\rangle}=+{\bf e}_{z}$
along $x$ direction and ${\bf e}_{\langle\bm{r}\bm{r}'\rangle}=-{\bf e}_{z}$
along $y$ direction, respectively. The clean AM is described by an
effective Hamiltonian $H_{0}({\bf k})=-(2t\cos k_{x}+2t\cos k_{y}+\mu)-2t_{J}(\cos k_{x}-\cos k_{y})\sigma_{z}$,
where ${\bf k}=(k_{x},k_{y})$ is the momentum vector. Disorder is
introduced in the AM central region of the AMJJ via a random on-site
potential as

\begin{equation}
h_{\mathrm{dis}}=\sum_{\bm{r}\in\mathrm{AMs}}V_{{\bf r}\sigma}C^{\dagger}_{\bm{r}\sigma}C_{\bm{r}\sigma},
\end{equation}
where $V_{{\bf r}\sigma}$ is drawn randomly from the interval $[-W/2,W/2]$
with $W$ being the disorder strength. At the interface layers of
normal metals, the onsite potential takes a constant as $V_{{\bf r}\sigma}=V_{b}$,
while it vanishes in the superconducting leads.

The supercurrent across the junction is obtained using the standard
recursive Green's function as \citep{Furusaki94physica,Asano01prb,Asano06prl}

\begin{alignat}{1}
I(\phi)= & \frac{ieT}{\hbar}\sum_{\omega_{n}}\mathrm{Tr}\{\tau_{z}H_{j,j+1}\hat{G}_{j+1,j}(i\omega_{n})\nonumber \\
 & \ \ \ \ \ \ \ \ \ \ -\tau_{z}H_{j+1,j}\hat{G}_{j,j+1}(i\omega_{n})\},
\end{alignat}
where $T$ is the temperature, $H_{j,j+1}$ is the hopping matrix
between two adjacent $j$- and $(j+1)$-th lattice layers in the middle
junction part, and $\hat{G}_{j,j+1}(i\omega_{n})$ is the corresponding
Nambu Green's function evaluated at Matsubara frequency $\omega_{n}=(2n+1)\pi/\beta$
with $\beta=1/k_{B}T$. Here, $n$ is an integer number, $k_{B}$
is the Boltzmann constant, and $\tau_{z}$ the Pauli matrix in Nambu
space. The temperature dependence of superconducting gap is taken
as $\Delta(T)=\Delta_{0}\tanh[1.74\sqrt{T_{c}/T-1}]$, where $\Delta_{0}=1.76k_{B}T_{c}$
and $T_{c}$ is the critical temperature.

The symmetry properties of AMJJs impose strong constraint on the current-phase
relation (CPR). The AM breaks time-reversal symmetry $\mathcal{T}$
and four-fold rotation symmetry $C_{4}$, while it respects the combined
symmetry $C_{4}\mathcal{T}$ as $[C_{4}\mathcal{T}]H_{0}({\bf k})C_{4}\mathcal{T}]^{-1}=H_{0}(-{\bf k})$
with $C_{4}=e^{-i\frac{\pi}{4}\sigma_{z}}$ and $\mathcal{T}=i\sigma_{y}K$.
For AMJJs, this particular symmetry implies $[C_{4}\mathcal{T}]\mathcal{H}(\phi)[C_{4}\mathcal{T}]^{-1}=\mathcal{H}(-\phi)$
\citep{LuB24prl,ChengQ24prb}. Correspondingly, the free energy, i.e.
$F(\phi)=-\frac{\ln Z(\phi)}{\beta}$ with $Z(\phi)=\mathrm{Tr}[e^{-\beta\mathcal{H}(\phi)}]$,
follows a relation $F(\phi)=F(-\phi)$. The supercurrent can be derived
as $I(\phi)=\partial_{\phi}F(\phi)$, yielding a constrained CPR as
$I(\phi)=-I(-\phi)$. Therefore, the allowed phases in AMJJs are just
$0$ phase, $\pi$ phase, and $\varphi$ phase. As we show below,
the $0$ phase and $\pi$ phase are dominated by the first harmonic
in the CPR while the anomalous $\varphi$ phase is dominated by higher-order
harmonics, e.g. the second- and third-order harmonics. In the following,
we investigate disorder-induced phase transitions among these phases.

\begin{figure*}
\includegraphics[scale=0.15]{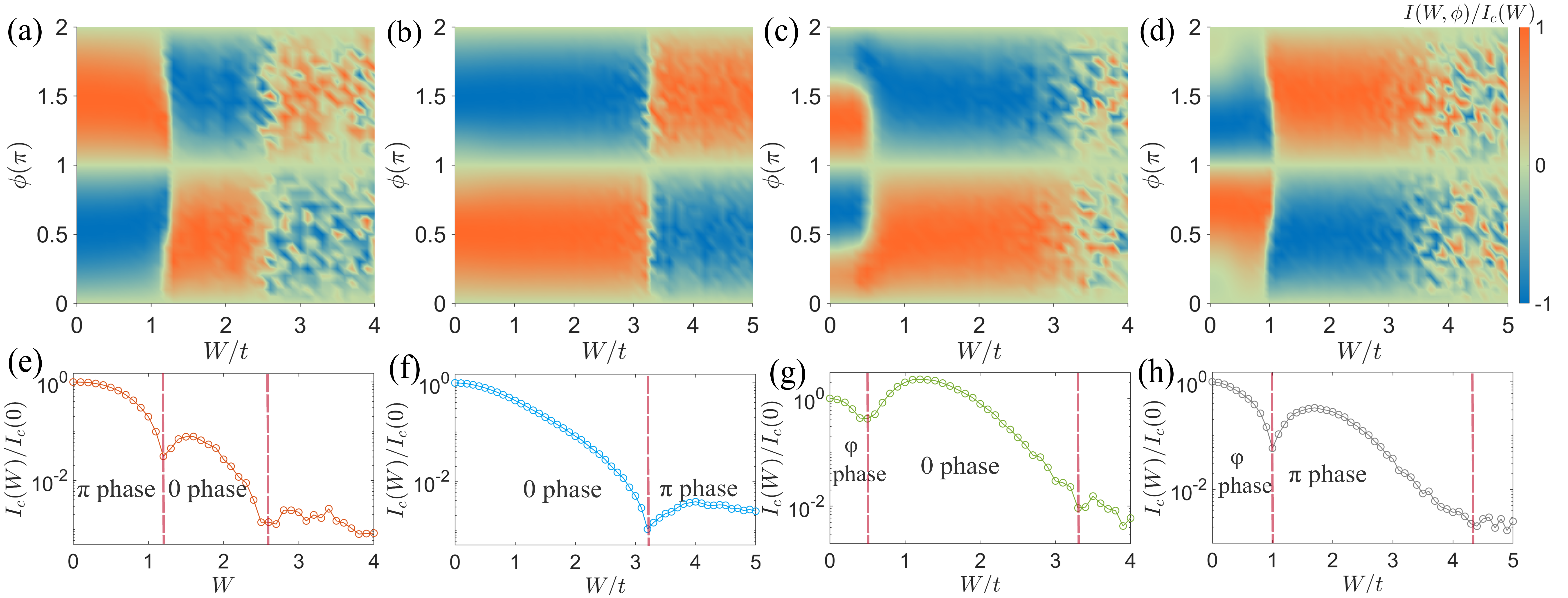}\caption{(a-d) Evolution of the CPR with increasing disorder strength $W$.
In the clean limit with $W=0t$, the AMJJ is in $\pi$ phase for (a)
with $L_{x}=16$, $0$ phase for (b) with $L_{x}=8$, $\varphi$ phase
in (c) with $L_{x}=18$, and $\varphi$ phase in (d) with $L_{x}=14$,
respectively. (e-h) The critical supercurrent as a function of $W$
corresponding to panels (a-d). Other parameters are $L_{y}=30,V_{b}=2t$,
$t_{J}=0.3t,\Delta=0.1t$, $T=0.05T_{c}$, and $\mu=-2t$. We have
taken $5\times10^{3}$ disorder configurations for all plots.\label{fig2:Phasetransition}}
\end{figure*}

\section{Results}

\subsection{Transitions between exotic $\pi$ phase and conventional $0$ phase}

We first examine the fate of exotic $\pi$ phase in presence of disorder.
In the clean phase, the CPR depends on the junction length of $L_{x}$
\citep{Ouassou23PRL,ZhangSB24NC}. With a proper junction length,
the interested $\pi$ phase can be realized. Specifically, the AMJJ
resides in the $\pi$ phase at $L_{x}=16$ in the clean limit under
our parameter settings \citep{Li2026SM}. Then we turn on the disorder
and study the evolution of the CPR. Figure \ref{fig2:Phasetransition}(a)
displays the normalized CPR $I(W,\phi)/I_{c}(W)$ as a function of
disorder strength $W$. With increasing $W$, the system undergoes
a transition to the conventional $0$ phase at a critical disorder
strength $W_{c1}\approx1.2t$. The corresponding critical current
is plotted in Fig. \ref{fig2:Phasetransition}(e). Across the transition,
the critical current $I_{c}$ experience a sharp dip and changes its
sign, consistent with the $\pi$ to 0 phase transition. These results
indicate that the exotic $\pi$ phase is not robust against disorder
whereas can be driven into a conventional $0$ phase even at a relatively
weak disorder strength. Note that the fluctuation under strong disorder
makes the critical current difficult to reach exactly zero at the
transition point.

With further increasing $W$, the supercurrent is strongly suppressed
{[}see Fig. \ref{fig2:Phasetransition}(e){]}. For $W_{c}>2.5t$,
the critical current is reduced to an order of $I_{c}(W)/I_{c}(0)\sim10^{-3}$
{[}Note that the critical current $I_{c}(0)$ is at the order of $0.01\frac{e\Delta}{\hbar}$
to $0.1\frac{e\Delta}{\hbar}$, depending on the initial conditions
such as the junction width and length{]}. In this regime, the strong
fluctuation of supercurrent due to disorder renders the CPR ill-defined
even after averaging over $5\times10^{3}$ configurations.\textcolor{blue}{{}
}In this limit, mesoscopic conductance fluctuations dominate in the
normal-state transport, which directly suppress the supercurrent.
We therefore exclude this strongly fluctuating regime from further
analysis. Similar strongly fluctuating regions are also observed in
other cases in Fig. \ref{fig2:Phasetransition}. 

Notably, disorder can also induce a transition from the conventional
$0$ phase to the exotic $\pi$ phase in a reversal direction. This
behavior is demonstrated in Figs. \ref{fig2:Phasetransition}(b) and
\ref{fig2:Phasetransition}(f). In the clean limit, the AMJJ resides
in the $0$ phase. As increasing $W$, the critical current decays
rapidly until the phase transition point at around $W_{c}\approx3.2t$.
Upon further increasing $W$, the supercurrent changes its sign, indicating
a transition into the $\pi$ phase. After the phase transition, the
critical current is reduced to an order of $I_{c}(W)/I_{c}(0)\sim10^{-2}$,
while the CPR remains well-defined.

As we have shown, the impact of disorder in AMJJs manifests in two
main aspects: (i) it drives exotic phase transitions; (ii) it suppresses
the critical supercurrent. We elucidate these two points as follows.
In AMJJs, the Cooper pairs tunneling through the junction will acquire
a finite momentum $\delta q\propto t_{J}$ even there is no net magnetization
\citep{Ouassou23PRL,ZhangSB24NC}. The supercurrent across the junction
of length $L_{x}$ is thus determined by an extra phase factor $\delta qL_{x}$
as $I\propto\cos(\delta qL_{x})$ \citep{Buzdin05rmp,ZhangSB24NC}.
Disorder modifies this behavior by effectively reducing the altermagnetic
spin-split strength, $t_{J}\to t_{J}'$, due to the smearing of the
anisotropic spin-dependent band structure \citep{LiCA26prl,ZengL25prb}.
This thus leads to a renormalized phase shift to $\delta q'L$. Therefore,
the critical current can flip its sign with changing the phase shift,
enabling the transition between the $0$ and $\pi$ phases. Note that
the scattering from disorder is detrimental to superconducting coherence
\citep{Furusaki94physica}. The strong disorder destroys the coherence
of tunneling Cooper pairs such that the critical current is substantially
suppressed, giving rise to the strongly fluctuating regime.

\subsection{Fragile $\varphi$ phase}

\begin{figure}
\includegraphics[width=1\linewidth]{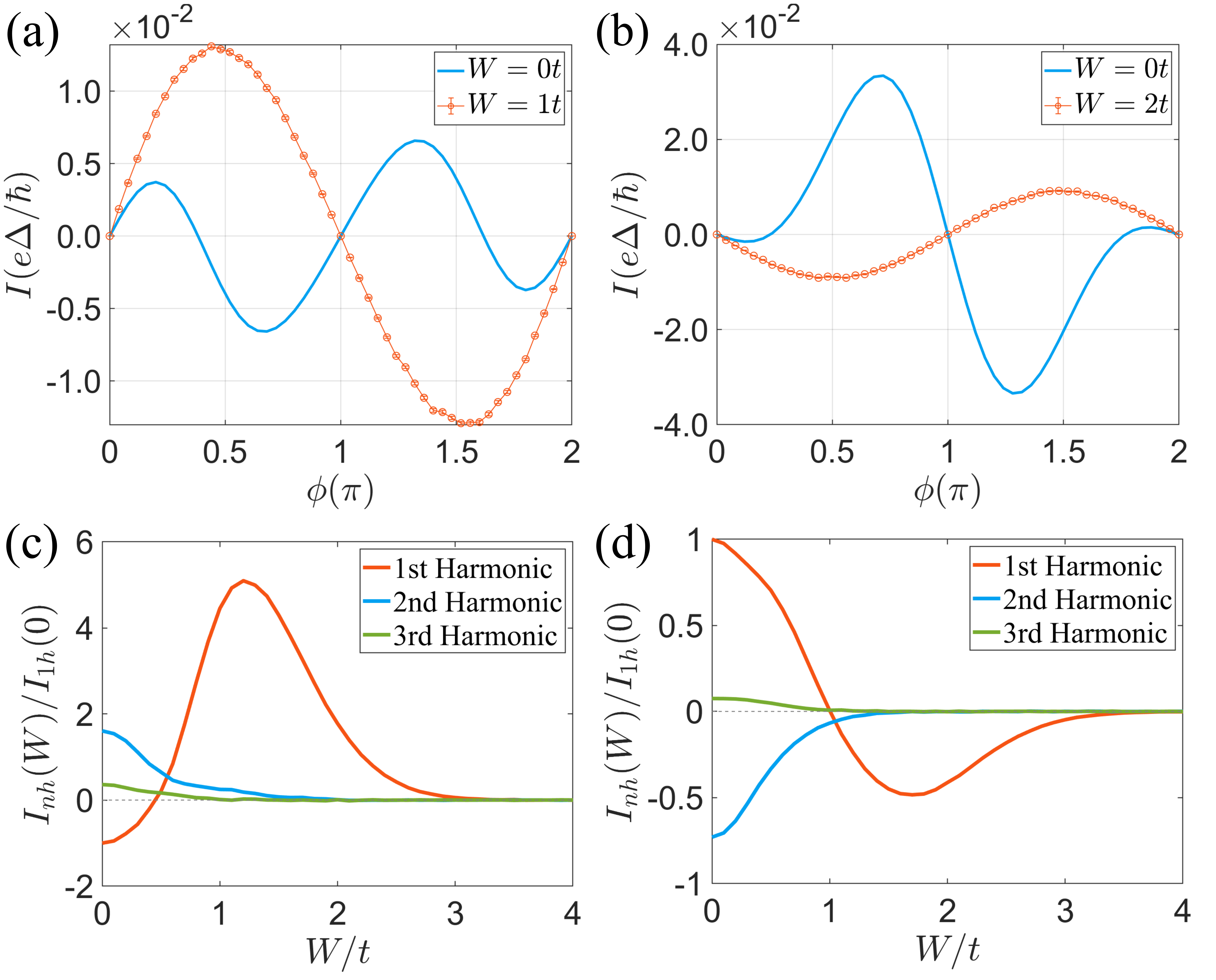}

\caption{(a-b) CPRs of the AMJJ at different disorder strength $W$. Panel
(a) demonstrates $\varphi$ phase before the transition and $0$ phase
after the transition. The AM junction length is of $L_{x}=18$. Panel
(b) demonstrates $\varphi$ phase before transition and $\pi$ phase
after transition. The AM junction length is of $L_{x}=14$. (c-d)
Corresponding harmonic components in CPRs as a function of $W$, corresponding
to Fig. \ref{fig2:Phasetransition}(c) and \ref{fig2:Phasetransition}(d),
respectively. Other parameters are the same as in Fig. \ref{fig2:Phasetransition}.
\label{fig:Harmonics}}
\end{figure}

AMJJs can host the anomalous $\varphi$ phase, which arises when higher-order
harmonics $I_{n}\sin(n\phi)$ (with $n\geq2$) dominates the CPR \citep{Ouassou23PRL,LuB24prl}.
In this case, the supercurrent can take zero values at $\phi\neq0,\pi$.
The $\varphi$ phase, however, turns out to be highly fragile under
disorder. We illustrate this result in Figs. \ref{fig2:Phasetransition}(c-d)
and \ref{fig2:Phasetransition}(g-h). For illustration, we choose
the systems initially in a $\varphi$ phase in the clean limit with
junction length $L_{x}=14$ and $L_{x}=18$, respectively. Upon introducing
disorders, even at weak strength, the $\varphi$ phase will be driven
into either the $0$ phase {[}Fig. \ref{fig2:Phasetransition}(c){]}
or the $\pi$ phase {[}Fig. \ref{fig2:Phasetransition}(d){]}, depending
on initial conditions. The critical current drops at the critical
points, signaling a phase transition. Surprisingly, the critical current
can even be slightly enhanced after the phase transition {[}Fig. \ref{fig2:Phasetransition}(g){]}.
The typical CPRs before and after phase transitions are presented
in Figs. \ref{fig:Harmonics}(a) and \ref{fig:Harmonics}(b), respectively.
Clearly, the induced $0$ phase and $\pi$ phase after phase transitions
exhibit smooth and sinusoidal CPRs.

The appearance of $\varphi$ phase is dominated by higher-order harmonics
in the CPRs \citep{LuB24prl}. Thus we explain the fragility of the
$\varphi$ phase in terms of suppression of higher-order harmonics
by disorder. To this end, we analyze the harmonic content via Fourier
decomposition of the CPR as
\begin{equation}
I(W,\phi)=\sum^{\infty}_{n=1}I_{nh}(W)\sin(n\phi),
\end{equation}
where $I_{nh}(W)$ is the amplitude of $n$-th order harmonic. We
perform fast Fourier transformation of CPRs corresponding to transitions
of $\varphi$ phase in Figs. \ref{fig2:Phasetransition}(c) and \ref{fig2:Phasetransition}(d).
We illustrate this point in Fig. \ref{fig:Harmonics}(c) as an example.
The three lowest orders with $n\leq3$ are plotted as a function of
disorder strength $W$. In the clean limit, the second harmonic $I_{2h}$
is comparable to first harmonic $I_{1h}$, while higher-order contributions
are negligible. As increasing the disorder strength $W$, $I_{2h}$
and $I_{3h}$ drop to zero in a monotonic way. In contrast, the first
harmonic $I_{1h}$ exhibits a non-monotonic behavior: its magnitude
decays to zero first, then changes sign, and subsequently increases
to a maximum before being totally suppressed by strong disorder. As
a result, the $\varphi$ phase collapses once the higher-order harmonics
are fully suppressed, leaving a CPR dominated by the first harmonic.
This behavior can also be understood from the free energy point of
view. Note that the CPR can be derived from $I(\phi)=\partial_{\phi}F(\phi)$.
The $\varphi$ phase corresponds to local extrema of free energy function
$F(\phi)$ at $\phi\neq0,\pi$ \citep{LuB24prl}. These local extrema
are not protected by any symmetries, unlike the global ones at $\phi=0/\pi$.
Random potentials smear out these features, stabilizing only the global
extrema at $\phi=0/\pi$, thereby suppressing higher-order harmonics
and destroying the $\varphi$ phase. Consequently, disorder cannot
induce a transition from the $\pi$ phase and $0$ phase back to the
$\varphi$ phase in a reversal direction, consistent with the results
presented in Fig. \ref{fig2:Phasetransition}. Note that the sign
of the first harmonic determines whether the $\varphi$ phase is driven
to a $0$ phase or a $\pi$ phase.

\subsection{Dependence on junction width and temperature}

\begin{figure}
\includegraphics[width=1\linewidth]{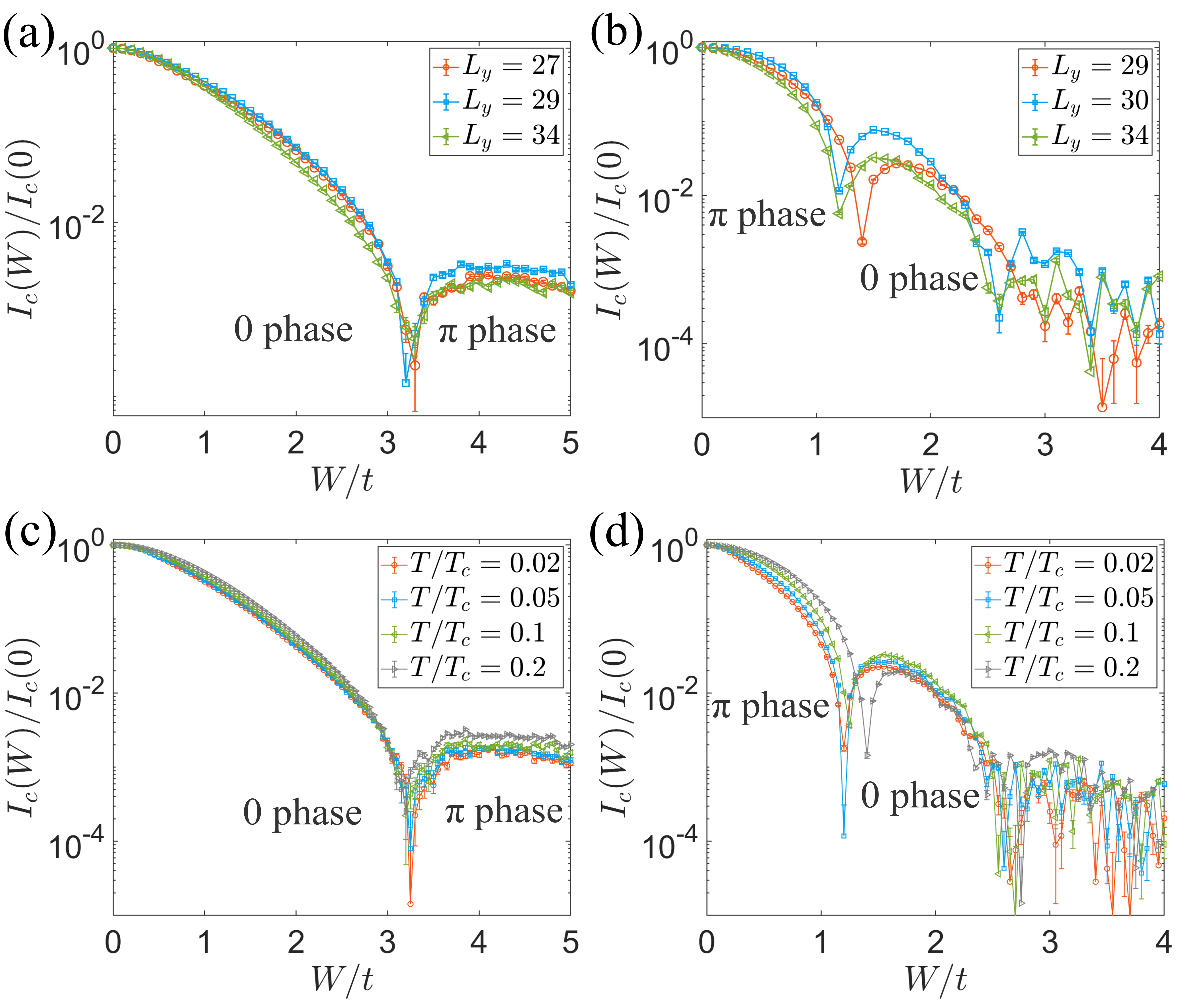}

\caption{(a,b) The width dependence of critical supercurrent as a function
of disorder strength $W$. The temperature is fixed at $T=0.05T_{c}$
and the junction length is $L_{x}=8$. (c,d) The temperature dependence
of critical supercurrent as a function of disorder strength $W$.
The width of the junction is fixed at $L_{y}=34$ and the junction
length is $L_{x}=16$. We have taken $1\times10^{4}$ disorder configurations
for all plots. Other parameters are $V_{b}=2t$, $t_{J}=0.3t,\Delta=0.1t$,
and $\mu=-2t$. \label{fig:Temp}}
\end{figure}

To demonstrate the generality of disorder-induced phase transitions
in AMJJs, we examine the dependence on junction width and temperature
\citep{Heras26prb}. We illustrate this point by taking the transitions
between $0$ phase and $\pi$ phase as an example, as shown in Fig.
\ref{fig:Temp}. Explicitly, Figs. \ref{fig:Temp}(a) and \ref{fig:Temp}(b)
display the critical current $I_{c}(W)/I_{c}(0)$ as a function of
$W$ for different junction width $L_{y}$. The transitions between
the $0$ phase and $\pi$ phase show robustness against small changes
of the width $L_{y}$, whereas the critical disorder exhibits slight
shift. The temperature dependence is presented in Figs. \ref{fig:Temp}(c)
and \ref{fig:Temp}(d). Note that we follow the same form of $\Delta(T)=\Delta_{0}\tanh[1.74\sqrt{T_{c}/T-1}]$
as before. With increasing $W$, the critical current decays rapidly
prior to the phase transition. This decay is more pronounced at lower
temperatures, indicating the stronger influence of disorder effect
on the superconducting transport. The phase transition persists at
different temperatures, while the corresponding critical disorder
strength is slightly modified. At strong disorder regime in Figs.
\ref{fig:Temp}(b) and \ref{fig:Temp}(d), the critical current demonstrates
strong fluctuations corresponding Fig. \ref{fig2:Phasetransition}(a).

\subsection{$d_{xy}$-wave case}

\begin{figure}
\includegraphics[width=1\linewidth]{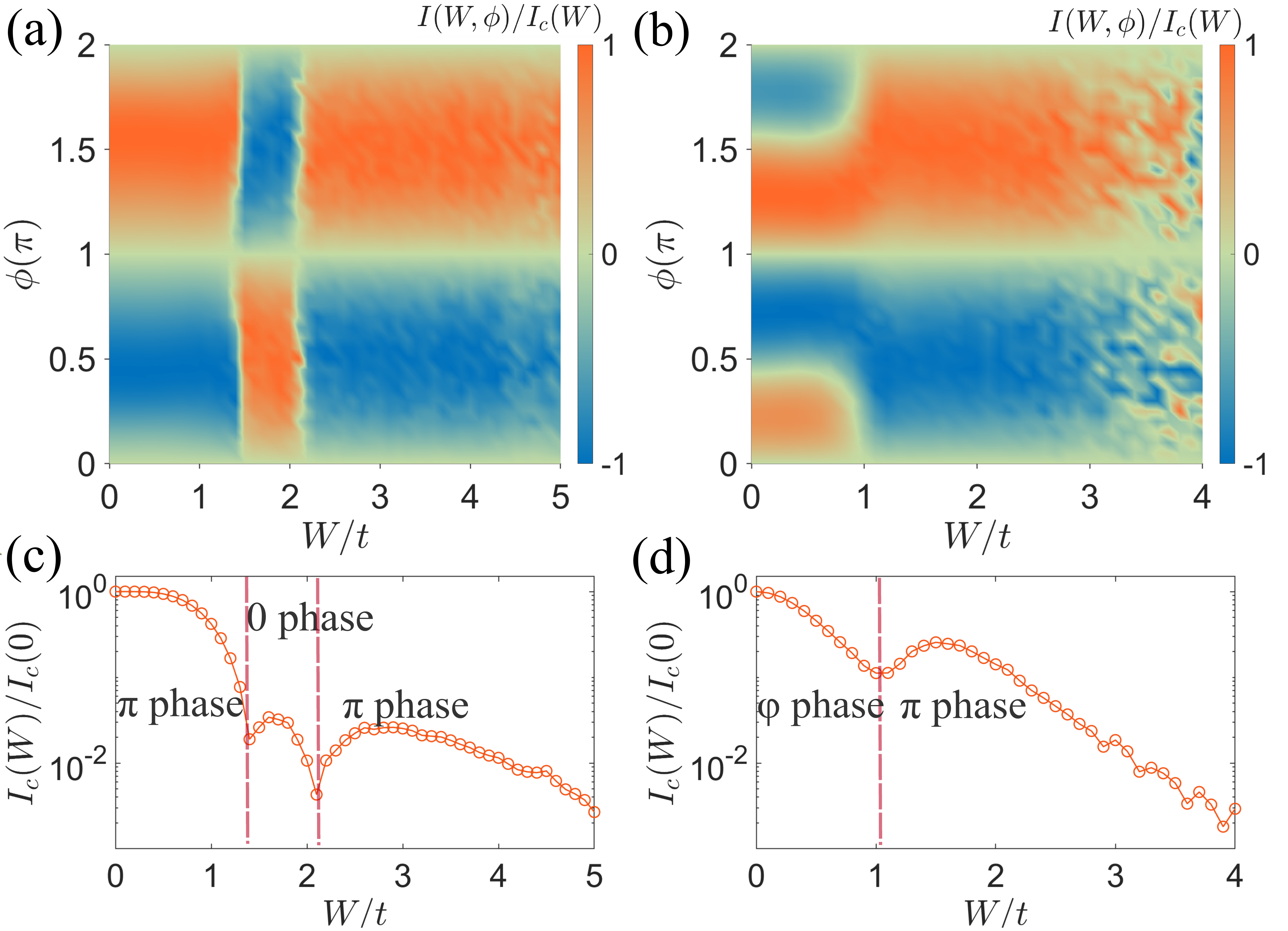}

\caption{(a-b) Evolution of the CPR with increasing disorder strength $W$
for $d_{xy}$-wave AMs. In the clean limit, the AMJJ is in $\pi$
phase for (a) with $L_{x}=12$, and $\varphi$ phase for (b) with
$L_{x}=22$. (c-d) The critical supercurrent as a function of $W$,
corresponding to panels (a-b). Other parameters are $L_{y}=30,V_{b}=2t$,
$t_{J}=0.3t,\Delta=0.1t$, $T=0.05T_{c}$, and $\mu=-2t$. We have
taken $5\times10^{3}$ disorder configurations for all plots. \label{fig:dxy-wave}}
\end{figure}

The anisotropy of AM orders has been found to affect the Josephson
effect significantly \citep{Ouassou23PRL,Beenakker23prb,ZhangSB24NC,ChengQ24prb}.
Therefore, we consider disorder-induced phase transitions in AMJJs
with $d_{xy}$-wave altermagnets, as a comparison to the $d_{x^{2}-y^{2}}$-wave
case. In the clean limit, the lattice model can be obtained by discretizing
the effective Hamiltonian \citep{ChengQ24prb} $H_{1}({\bf k})=-[t(k^{2}_{x}-k^{2}_{y})+\mu]\sigma_{0}+2t_{J}k_{x}k_{y}\sigma_{z}$.
The general altermagnetic order comes in the form $-t_{J}[(k^{2}_{x}-k^{2}_{y})\cos(2\alpha)-2k_{x}k_{y}\sin(2\alpha)]\sigma_{z}$,
where the angle $\alpha$ determines the AM order orientation. For
instance, the typical $d_{x^{2}-y^{2}}$-wave AM takes $\alpha=0$
and $d_{xy}$-wave AM takes $\alpha=\pi/4$. The random on-site potential
$V_{{\bf r}\sigma}$ is added in the altermagnetic junction as described
above. Figure \ref{fig:dxy-wave} presents the main results for disorder-induced
phase transitions in AMJJs with $d_{xy}$-wave altermagnets. As shown
in Figs. \ref{fig:dxy-wave}(a) and \ref{fig:dxy-wave}(c), an initial
$\pi$ phase is first driven to a $0$ phase at a critical disorder
strength $W_{c}\approx1.3t$. Interestingly, upon further increasing
$W$, the system reenters to $\pi$ phase again after passing through
a narrow window of $0$ phase, demonstrating a particular re-entrant
phenomenon. In the $d_{xy}$-wave case, the $\varphi$ phase remains
fragile under disorder. As shown in Fig. \ref{fig:dxy-wave}(b), the
$\varphi$ phase undergoes a transition to $\pi$ phase at a critical
point $W_{c}\approx1.0t$. The corresponding critical current curve
has a minimum around the critical disorder strength as shown in Fig.
\ref{fig:dxy-wave}(d). These consistent results confirm that disorder-induced
phase transitions are generic features in the AMJJs even with different
AM order orientations.

\section{Discussion}

The disorder-induced phase transitions of AMJJs are general phenomena
even considering different types of disorder such as magnetic disorder.
The magnetic disorder takes the form $V_{{\bf r}}=\sum_{s=0,x,y}w^{s}_{{\bf r}}\sigma_{s}$,
with spin-dependent random potentials at each lattice site. Here,
the $w^{0}_{{\bf r}}$ term indicates the usual nonmagnetic type disorder,
and the $w^{x}_{{\bf r}}\sigma_{x}$ and $w^{y}_{{\bf r}}\sigma_{y}$
terms are magnetic disorder that couples spin degrees of freedom in
AMs. Under the magnetic disorder, transitions between the $\pi$ phase
and the $0$ phase also occur and the fragility of $\varphi$ phase
remains \citep{Li2026SM}. In contrast, when the $d$-wave AM is replaced
by odd-parity $p$-wave unconventional magnets, no phase transitions
are observed \citep{Li2026SM}. It is interesting to further explore
impact of the pairing amplitude symmetries in disordered AMJJs, such
as the odd-parity pairing. We employ dimensionless model parameters
in our calculation to describe a generic AMJJ. While the model parameters
are not fitted to a specific material, the choice $t_{J}=0.3t$ represents
the reasonable fraction of the electronic hopping scale in altermagnets,
such as CrSb and $\mathrm{KV_{2}Se_{2}O}$ where altermagnetic band
splittings are of several hundred meV to approximately 1 eV.

Our results should be directly relevant to experimental realization
and potential functionalities of AMJJs. For instance, in the confirmed
metallic altermagnets materials $\mathrm{KV_{2}Se_{2}O}$ and $\mathrm{Rb_{1-\delta}V_{2}Te_{2}O}$
with defects or high doping\textcolor{black}{{} \citep{JiangB25NP,FYZhang25NP,DingJY24PRL,ZengM24AS,YangGW25NC},}
disorder will substantially modify their transport features \citep{LiCA26prl,ZengL25prb,Heras26prb}.
Henceforth, the disorder effect should be considered in the fabrication
of related AMJJs. 

\section{Conclusion}

In summary, we have demonstrated that disorder in AMJJs plays a crucial
role: it induces phase transitions between distinct phases. In particular,
disorder drives phase transition from the exotic $\pi$ phase to conventional
$0$ phase even at a weak disorder strength, while the $0$ to $\pi$
phase transition is also enabled, establishing a reciprocal manner.
Furthermore, we reveal that the anomalous $\varphi$ phase is intrinsically
fragile in presence of disorder. It is driven to either a $0$ phase
or $\pi$ phase, depending on sign of the first harmonic component
of the CPR. Our findings reveal the pronounced sensitivity of AMJJs
to disorder such that the disorder effect should be considered for
further functionalities of AMJJs. 

\section{Acknowledgments}

The author acknowledges Bo Fu, Lun-Hui Hu, Huaiming Guo, Song-Bo Zhang,
and Bj\"orn Trauzettel for insightful discussions, and thanks B. Lu
and Y. Tanaka for helpful communications. This work was supported
by the start-up funding at HFNL (Grant No. QD2022600001) and W\"urzburg-Dresden
Cluster of Excellence ctd.qmat, EXC 2147.
%\bibliographystyle{apsrev4-1-etal-title-url-first}
%\bibliography{Refsdata}
%merlin.mbs apsrev4-1.bst 2010-07-25 4.21a (PWD, AO, DPC) hacked
%Control: key (0)
%Control: author (72) initials jnrlst
%Control: editor formatted (1) identically to author
%Control: production of article title (1) required
%Control: page (0) single
%Control: year (1) truncated
%Control: production of eprint (0) enabled
%

\appendix
\numberwithin{equation}{section}\setcounter{figure}{0}\global\long\def\thefigure{S\arabic{figure}}
\global\long\def\thesection{S\arabic{section}}
\global\long\def\thesubsection{\Alph{subsection}}

\begin{widetext}
\begin{center}
\textbf{\large{}Supplemental materials of ``Phase Transitions in Disordered Altermagnetic Josephson Junctions''}{\large{} }
\par\end{center}{\large \par}
%\tableofcontents
\section{Evolution of current-phase relations under disorder}

In this section, we present the current-phase relations (CPRs) of
the disordered AMJJs before and after different types of phase transitions,
corresponding to Fig. \textcolor{blue}{2} in the main text. In the
clean limit, the CPRs change as a function of the junction length
$L_{x}$ with a fixed width $L_{y}$. As shown in Fig. \ref{fig:differentphase}(a),
the AMJJ exhibits the conventional $0$ phase, the exotic $\pi$ phase,
and the anomalous $\varphi$ phase at different junction lengths.
Upon introducing disorder, the the system undergoes phase transitions
that modify the form of the CPR. Figures \ref{fig:differentphase}(b)-\ref{fig:differentphase}(d)
demonstrate representative evolution of the CPR across such transitions.
As demonstrated in Fig. \ref{fig:differentphase}(b), the system is
in $\pi$ phase at the clean limit with $W=0$. With increasing the
disorder strength $W$, the system is driven to the conventional $0$
phase at a critical disorder strength $W\approx1.2t$. Note that the
CPR at $W=1.5t$ clearly corresponds to the $0$ phase.

For the $\varphi$ phase, disorder can also induce a transition into
the conventional $0$ phase, as illustrated in Fig. \ref{fig:differentphase}(c).
At weak disorder strength $W=1.0t$, the AMJJ is already locating
in a $0$ phase and the critical current is even enhanced compared
to the clean limit. Increasing $W$ further, the critical current
will be gradually suppressed as expected. Alternatively, the $\varphi$
phase can be driven to $\pi$ phase with different initial conditions,
as shown in Fig. \ref{fig:differentphase}(d). In this case, the system
experiences phase transition at around $W\approx1.0t$ and it is well
within the $\pi$ phase for $W=2.0t$. These results indicate that
the anomalous $\varphi$ phase is fragile in presence of disorder.

\begin{figure}
\centering

\includegraphics[clip,width=0.8\columnwidth]{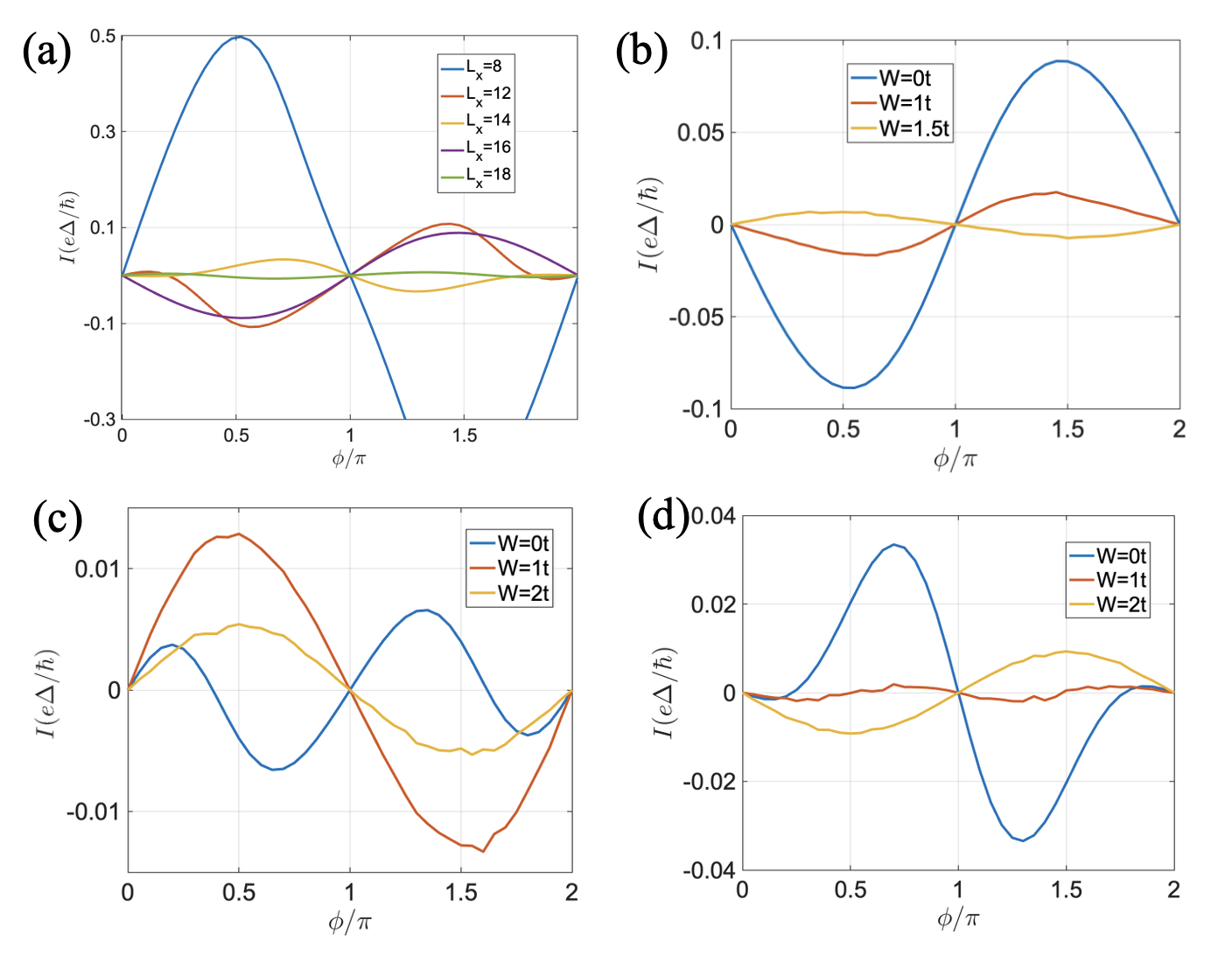}

\caption{(a) Plots of CPRs for different junction lengths. (b) Phase transitions
from the exotic $\pi$ phase to the conventional $0$ phase with increasing
disorder strength $W$. The junction length is fixed at $L_{x}=16$.
(c) Phase transitions from the anomalous $\varphi$ phase to the conventional
$0$ phase as increasing $W$. The junction length is fixed at $L_{x}=18$.
(d) Phase transitions from the $\varphi$ phase to the $\pi$ phase
as increasing $W$. The junction length is fixed at $L_{x}=14$. Other
parameters are $L_{y}=30,V_{b}=2t$, $t_{J}=0.3t,\Delta=0.1t$, $T=0.05T_{c}$,
and $\mu=-2t$. We have taken $5\times10^{3}$ disorder configurations
for all plots.}

\label{fig:differentphase}
\end{figure}

\section{Evolution of different order of harmonics during the 0-$\pi$ phase
transition}

In this section, we analyze the evolution of harmonics components
during the $0$-$\pi$ phase transition, complementing the results
presented in Fig. \textcolor{blue}{3} of the main text. As we have
shown, the CRPs at different disorder strength are generally expressed
as $I(W,\phi)=\sum_{n=1}^{\infty}I_{nh}(W)\sin(n\phi)$ where $I_{nh}(W)$
denotes the amplitude of $n$-th order of harmonics. To extract the
value of $I_{nh}(W)$, we perform Fourier transformation for the corresponding
CPRs. We first focus on the phase transition from the exotic $\pi$
phase to the $0$ phase, as shown in Fig. \ref{fig:0piHarmonics}
(a) {[}corresponding to Figs. \textcolor{blue}{2}(a) and \textcolor{blue}{2}(e)
of the main text{]}. In the $\pi$ phase, the CPR is dominated by
the first harmonic $I_{1h}$ with a negative value. The higher-order
ones, such as the second and third orders, are vanishingly small compared
to $I_{1h}$. With increasing disorder strength $W$, the higher-order
harmonics $I_{2h}(W)$ and $I_{3h}(W)$ rapidly decay to zero. In
contrast, the first harmonic $I_{1h}(W)$ (the absolute value) drops
gradually and changes sign across the phase transition point, signaling
the transition from the $\pi$ phase to the $0$ phase. Increasing
$W$ further, the first harmonic of $I_{1h}(W)$ will finally be suppressed
close to zero, consistent with the substantially reduced critical
current at the strong disorder strength.

A similar result is observed for the reversal transition from the
convectional $0$ phase to the exotic $\pi$ phase {[}Fig. \ref{fig:0piHarmonics}
(b){]}. In the clean limit, the CPR is dominated by the first harmonic
$I_{1h}$ and the higher-order harmonic values $I_{2h}$ and $I_{3h}$
are vanishingly small. As increasing $W$, the first harmonic $I_{1h}$
decays gradually and changes its sign across the phase transition
point {[}see the zoomed region in Fig. \ref{fig:0piHarmonics}(b){]}.
The system is transited to a $\pi$ phase afterwards with significantly
reduced critical current.

\begin{figure}
\centering

\includegraphics[clip,width=0.9\columnwidth]{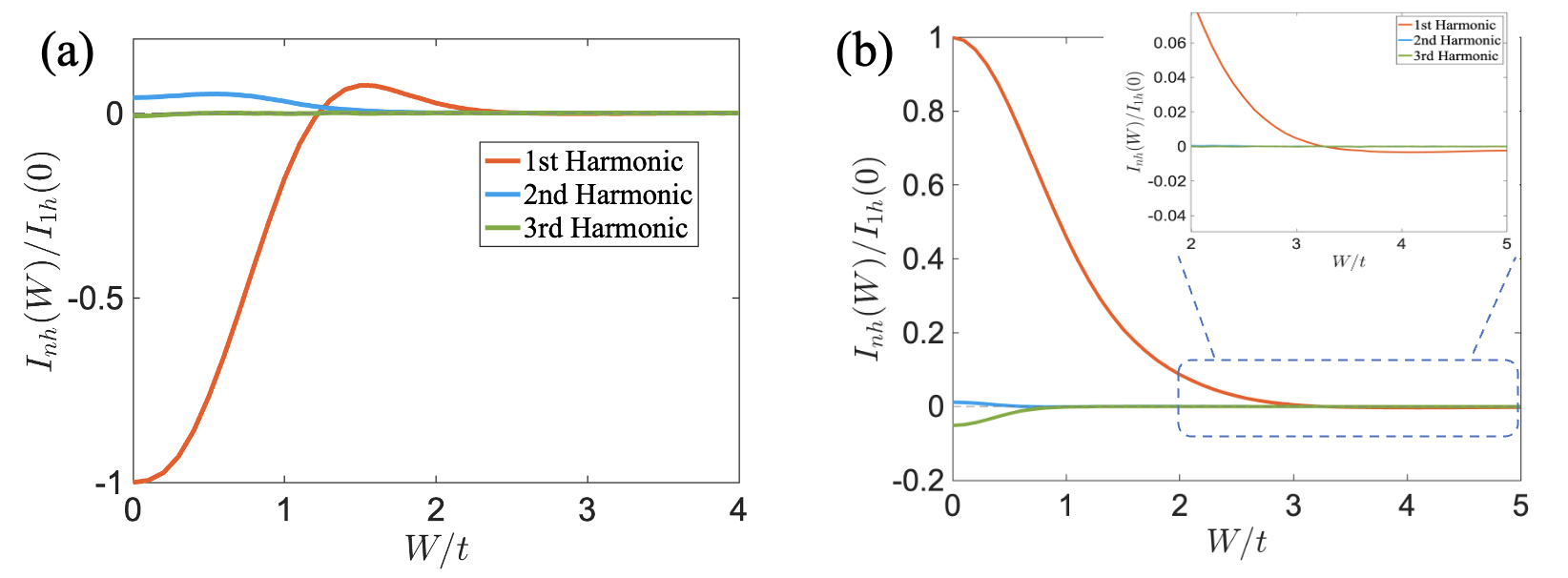}

\caption{(a) Different orders of harmonics in the CPRs as a function of disorder
strength $W$, corresponding to phase transitions from the $\pi$
phase to the $0$ phase. (b) Similar to (a) but with the phase transitions
from the $0$ phase to the $\pi$ phase. The zoomed inset plots the
sign change of $I_{1h}(W)$. The data of panels (a) and (b) are from
Figs. \textcolor{blue}{2}(a) and \textcolor{blue}{2}(b) of the main
text, respectively.}

\label{fig:0piHarmonics}
\end{figure}

\section{General properties of AMJJs under general magnetic disorder}

In this section, we demonstrate the disorder-induced phase transitions
occur when more general type disorder is considered. To this end,
we take the magnetic disorder of the form $V({\bf r})=\sum_{s=0,x,y}w_{{\bf r}}^{s}\sigma_{s}$
with spin-dependent random potentials at each lattice site ${\bf r}$.
The random potential $w_{{\bf r}}^{s}$ takes value independently
from the uniformly distributed interval $[-W/2,W/2]$ with $W$ being
the disorder strength. Here, the $w_{{\bf r}}^{0}\sigma_{0}$ term
indicates the usual nonmagnetic type disorder, and the $w_{{\bf r}}^{x}\sigma_{x}$
and $w_{{\bf r}}^{y}\sigma_{y}$ terms denote magnetic disorder that
couples to the spin degrees of freedom. The main results are presented
in Fig. \ref{fig:magnetic disorder}. In Fig. \ref{fig:magnetic disorder}(a),
the initial $\pi$ phase is driven to the $0$ phase at a relatively
weak disorder strength. Compared to the non-magnetic disorder case
presented in Fig. \ref{fig:differentphase}(b), the critical disorder
strength is reduced: the system is already in the $0$ phase with
$W=1.0t$, whereas it remains in the $\pi$ phase at the same disorder
strength.

Figures \ref{fig:magnetic disorder}(b) and \ref{fig:magnetic disorder}(c)
further present the transitions from $\varphi$ phase to $0$ phase
and $\pi$ phase, respectively. The overall behavior is qualitatively
consistent with the non-magnetic disorder case presented in Figs.
\ref{fig:differentphase}(c) and \ref{fig:differentphase}(d), but
the values of critical disorder strength become smaller. This result
is reasonable since extra randomness is introduced in the spin degrees
of freedom, leading to a stronger decoherence effect.

\begin{figure}
\centering

\includegraphics[clip,width=0.9\columnwidth]{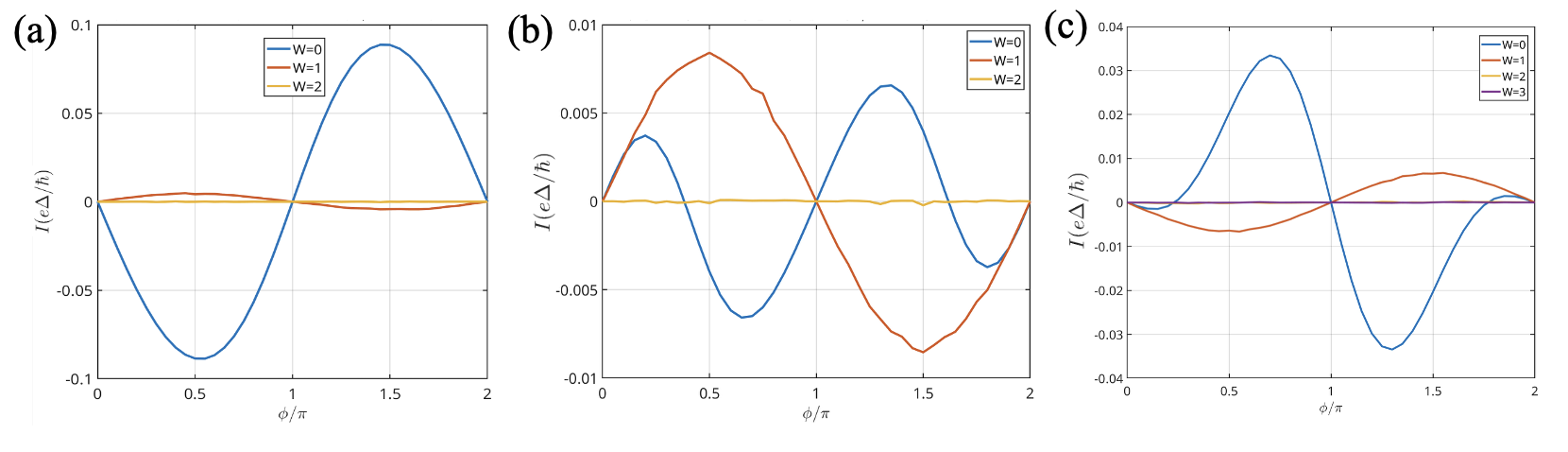}

\caption{Phase transitions of AMJJs under magnetic disorder. (a) Evolution
of CPRs demonstrates phase transitions from the $\pi$ phase to the
$0$ phase as increasing $W$. The junction length is fixed at $L_{x}=16$.
(b) Evolution of CPRs demonstrates phase transitions from the $\varphi$
phase to the $0$ phase. The junction length is fixed at $L_{x}=18$.
(c) Evolution of CPRs demonstrates phase transitions from the $\varphi$
phase to the $\pi$ phase. The junction length is fixed at $L_{x}=14$.
Other parameters are set at $L_{y}=30,V_{b}=2t$, $t_{J}=0.3t,\Delta=0.1t$,
$T=0.05T_{c}$, and $\mu=-2t$, the same as in the main text. We have
taken $5\times10^{3}$ disorder configurations for all plots.}

\label{fig:magnetic disorder}
\end{figure}

\section{Properties of disordered $p$-wave magnets Josephson junctions}

In this section, we investigate the disordered $p$-wave magnets Josephson
junctions. To this end, we employ the effective model for $p$-wave
magnets as \citep{hellenes24arXiv,Yamada25Nature}
\begin{equation}
H_{p}({\bf k})=t{\bf k}^{2}\sigma_{0}+pk_{x}\sigma_{x},
\end{equation}
where $t$ is the hopping amplitude and $p$ is the magnitude of spin
splitting. Note that the $p$-wave magnets do not break time reversal
symmetry, and it has momentum-resolved spin polarization $\langle S_{x}\rangle$.
Due to the odd-parity spin splitting, the system does not have inversion
symmetry. To construct the corresponding tight-binding Hamiltonian
on a 2D square lattice, we use the long-wavelength approximation that
$k_{x}\approx\sin(k_{x})$, $k_{x}^{2}\approx2(1-\cos k_{x})$. The
junction geometry is the same as in Fig. \textcolor{blue}{1}(a) of
the main text, with the $d$-wave AM being replaced by $p$-wave magnets.

The $p$-wave magnets Josephson junctions show similar properties
to the conventional Josephson junctions \citep{Fukaya25JPCM,Fukaya25prb},
i.e., the $0$ phase with sinusoidal CPRs. As shown in Fig. \ref{fig:p-wave Magnets}(a),
the CPR remains in the 0 phase for all junction lengths considered.
As increasing the junction length $L_{x}$, the critical current decays
gradually in an oscillating way, which is different from the conventional
Josephson junctions. We next introduce the disorder of type $V({\bf r})=w_{{\bf r}}^{0}\sigma_{0}$
at each lattice site, where $w_{{\bf r}}^{0}$ is drown from the uniformly
distributed interval $[-W/2,W/2]$, with $W$ denoting the disorder
strength. As the disorder strength increases, the $p$-wave magnets
Josephson junction remains in the $0$ phase {[}see Fig. \ref{fig:p-wave Magnets}(b){]}.
Correspondingly, the critical current is reduced by disorder {[}see
Fig. \ref{fig:p-wave Magnets}(c){]}. In contrast to the $d$-wave
AMJJs, no disorder-induced phase transition is observed.

\begin{figure}
\centering

\includegraphics[clip,width=0.9\columnwidth]{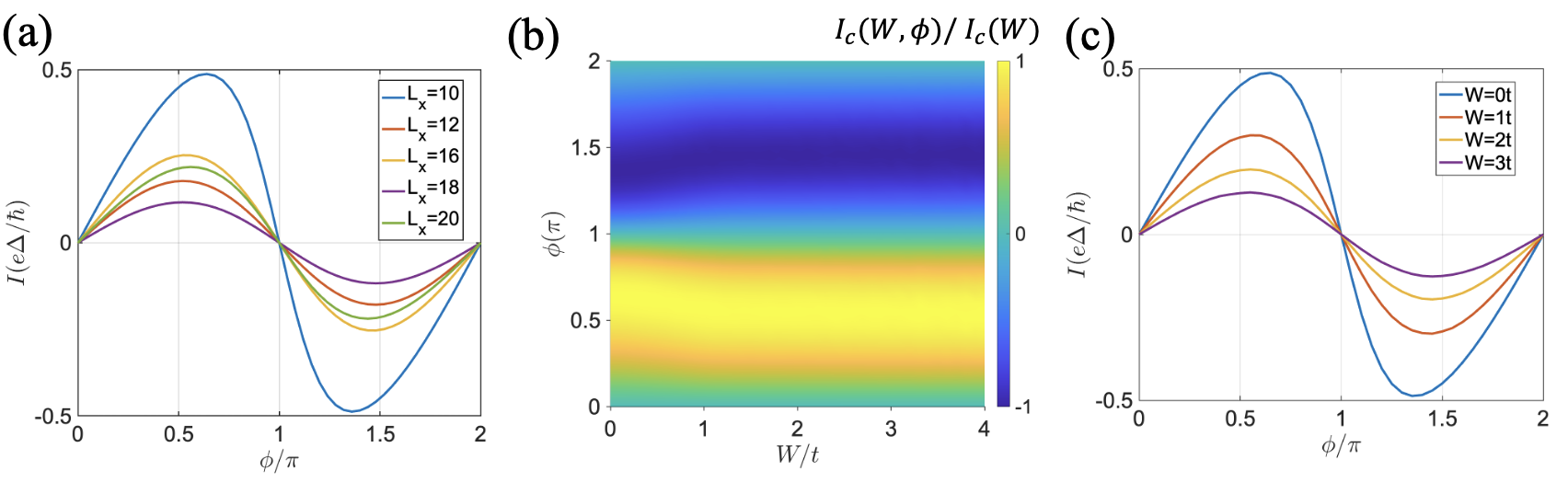}

\caption{(a) The CPRs in the $p$-wave magnets Josephson junctions as changing
the junction length $L_{x}$. (b) The evolution of CPRs as a function
of disorder strength $W$. Here we take $L_{x}=10$. (c) The corresponding
CPRs at different disorder strengths with a fixed junction length
$L_{x}=10$. Other parameters are set at $L_{y}=30,V_{b}=2t$, $p=0.3t,\Delta=0.1t$,
$T=0.05T_{c}$, and $\mu=-2t$. We have taken $2\times10^{3}$ disorder
configurations for panels (b) and (c).}

\label{fig:p-wave Magnets}
\end{figure}

\end{widetext}

\end{document}